\documentclass[aps,twocolumn,superscriptaddress]{revtex4}
\usepackage{amsmath}
\usepackage[colorlinks]{hyperref}
\usepackage{amssymb}
\usepackage{color}
\usepackage{graphicx,amsmath}
\hypersetup{colorlinks,citecolor=red,linkcolor=blue,urlcolor=blue}

\begin{document}
\title{Quantum Phase Transition in $\rm CeCoIn_5$:
Experimental Facts and Theory}
\author{V. R. Shaginyan}\email{vrshag@thd.pnpi.spb.ru} \affiliation{Petersburg
Nuclear Physics Institute of NRC "Kurchatov Institute", Gatchina,
188300, Russia}\affiliation{Clark Atlanta University, Atlanta, GA
30314, USA} \author{A. Z. Msezane}\affiliation{Clark Atlanta
University, Atlanta, GA 30314, USA} \author{M.~V. Zverev}
\affiliation{NRC Kurchatov Institute, Moscow, 123182, Russia}
\affiliation{Moscow Institute of Physics and Technology,
Dolgoprudny, Moscow District 141700, Russia}\author{Y. S. Leevik}
\affiliation{National Research University Higher School of
Economics, St.Petersburg, 194100, Russia}

\begin{abstract}
Condensed-matter community is involved in hot debate on the nature
of quantum critical points (QCP) governing the low-temperature
properties of heavy fermion metals. The smeared jump like behavior
revealed both in the residual resistivity $\rho_0$ and the Hall
resistivity $R_H$, along with the violation of the time invariance
symmetry $\mathcal{T}$ and the charge invariance $\mathcal{C}$,
including the violation of quasiparticle-hole symmetry, and
providing vital clues on the origin of both the non-Fermi-liquid
behavior and QCP. For the first time, based on a number of
important experimental data, we show that these experimental
observations point out unambiguously that QCP of $\rm CeCoIn_5$ is
accompanied by the symmetry violation, and QCP itself is
represented by the topological fermion-condensation quantum phase
transition (FCQPT) connecting two Fermi surfaces of different
topological charges.
\end{abstract}

\maketitle

%\end{document}

%\section{Introduction}

It is accepted that the fundamental physics of heavy fermion (HF)
metals is controlled by quantum phase transitions, see e. g.
\cite{ks,phys_rep94,vol,hall22,voj,phys_rep,book_20,Khod_2020,kir,tau}.
A quantum phase transition could be driven by control parameters
such as pressure $P$, number density $x$ of electrons (holes),
magnetic field $B$, etc, taking place at its quantum critical point
(QCP) at zero temperature, $T=0$. {The modern physics of
condensed-matter is represented by the experimental discovery of
flat bands since HF compounds with flat bands are numerous
\cite{catal}. As a result, one expects the existence of a general
physical mechanism generated by the presence of flat bands, forming
the universal properties of HF compounds.}

The magnetic field dependence of the Hall coefficient $R_H(B)$
provides information about QCP, determining the properties of HF
metals. Experiments have shown that the Hall coefficient $R_H(B)$
in the antiferromagnetic HF metal $\rm YbRh_2Si_2$ in magnetic
fields $B$ undergoes a jump in the zero temperature limit under the
application of $B$ tuning the metal from antiferromagnetic to a
paramagnetic state at $B=B_{c0}$ \cite{pash}. The jump takes place
when $B$ reaches the critical value $B_{c0}$ at which the N\'eel
temperature $T_N(B)$ of the antiferromagnetic transition vanishes,
$T_N(B\to B_{c0})\to0$. The jump is interpreted as a collapse of
the large Fermi surface precisely at QCP \cite{pash}. The fermion
condensation (FC) theory successfully explains such types of
behavior as both the universal scaling behavior and the transition
from the non-Fermi liquid (NFL) state to the Landau Fermi liquid
(LFL) one \cite{lanl1,phys_rep,book_20}. Moreover, the FC state is
characterized by broken $\mathcal{T}$ and $\mathcal{C}$ symmetries
\cite{phys_rep,phys_rep94,book_20,ph_scr,symm,atom} taking place at
the interaction driven topological fermion condensation phase
transition (FCQPT) that connects two Fermi surfaces of different
topological charges and forms flat bands \cite{ks,vol,phys_rep}.
Recent measurements on the prototypical HF metal $\rm CeCoIn_5$ of
the Hall coefficient allow one to interpret the observed quantum
phase transition as a delocalization quantum phase transition
without symmetry breaking in $\rm CeCoIn_5$, that is characterized
by the delocalization of $f$-electrons in the transition that
connects two Fermi surfaces of different volumes \cite{hall22}.

In this letter, for the first time, based on a number of important
experimental data, we reveal the quantum phase transition, that
formes the properties of archetypical HF metal $\rm CeCoIn_5$,
being characterized by the presence of flat bands \cite{chen} and
related to the violation of symmetries. We show that the measured
thermodynamic and transport properties yield direct evidence for
the apparent broken symmetries like the violation of the
$\mathcal{T}$ and $\mathcal{C}$ symmetries taking place at the
topological FCQPT that occurs in both the archetypical HF metals
$\rm CeCoIn_5$ and $\rm YbRh_2Si_2$.

%\section{Schematic $T-B$ phase diagram of $\rm
%CeCoIn_5$}

\begin{figure}[!ht]
\begin{center}
\includegraphics [width=0.47\textwidth]{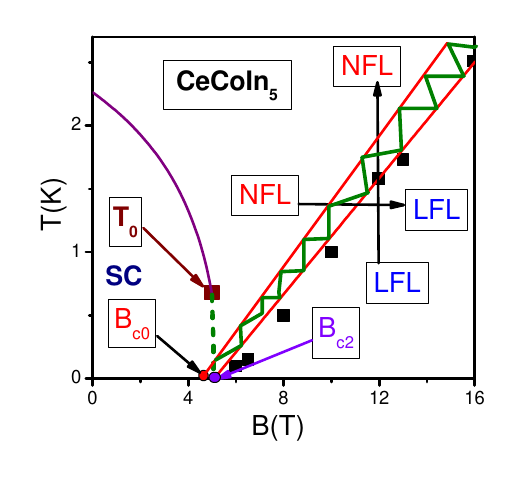}
\vspace*{-0.4cm}
\end{center}\vspace*{-0.8cm}
\caption{(color online) {Schematic $T-B$ phase diagram of $\rm
CeCoIn_5$, with the upper critical field $B_{c2}\simeq 5.1$ T along
the [001] direction.} The vertical and horizontal arrows crossing
the transition region are marked by the line depicting the LFL-NFL
and NFL-LFL transitions at fixed $B$ and $T$, respectively.
Experimental data obtained from resistivity $\rho(T)$ measurements
(upper bound of $T^2$ resistivity) are shown by the solid squares
\cite{pag1,pag2}. The hatched area indicates the crossover from the
LFL state, with  $\rho(T)\propto T^2$, to NFL one, with
$\rho(T)\propto T$. As shown by the solid curve, at $B<B_{c2}$ the
system is in its superconducting (SC) state, with $B_{c0}$ denoting
QCP hidden beneath the SC dome and shown by the red circle. The
superconducting critical field $B_{c2}\simeq 5.1$ T is depicted by
the violet circle. Superconducting-normal phase boundary
\cite{bianchi} is depicted by the solid and dashed curves. The
solid squares show the point at $T=T_0$ where the superconducting
phase transition $T_c$ changes from the second to the first order.}
\label{fig1}
\end{figure}

{Consider $T-B$ phase diagram of the HF metal $\rm CeCoIn_5$ with
the upper critical field $B_{c2}=5.1$ T along the [001] direction
\cite{pag1,pag2,bianchi,qcp11}. The application of magnetic field
along the [100] direction makes the upper critical field
$B_{c2}=11.8$ T and leads to the complicated superconducting part
of the phase diagram \cite{sc08,prl10}, that is not related with
our consideration of the phase diagram \ref{fig1} at relatively low
magnetic fields.} The phase diagram is presented schematically in
Fig.~\ref{fig1}. The HF metal $\rm CeCoIn_5$ is a superconductor
with $T_c=2.3$ K with the field tuned QCP occurring at the critical
field $B_{c0}\simeq5.0$ T, while the superconducting critical field
$B_{c2}\simeq 5.1$ T. Thus, QCP is hidden by the superconducting
state, as seen from Fig. \ref{fig1}. We note that $B_{c0}\simeq
B_{c2}$ is an accidental coincidence. Indeed, it is shown that the
application of increasing pressure $P$, QCP moves inside the
superconducting dome, thus removing the above
coincidence\cite{ronn}.

\begin{figure}
\begin{center}
\vspace*{-0.2cm}
\includegraphics [width=0.42\textwidth]{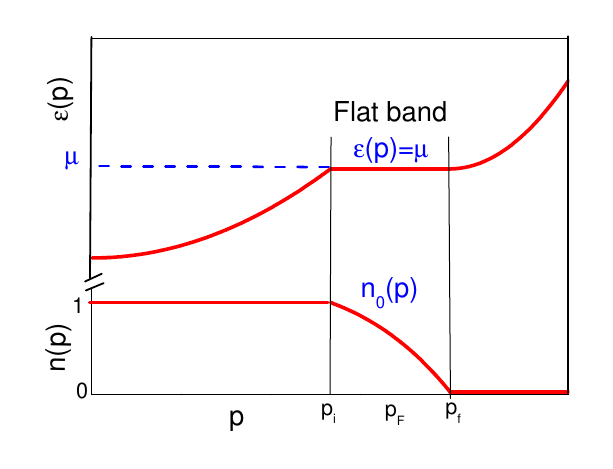}
\vspace*{-1.0cm}
\end{center}
\caption{(color online)  {The single-particle spectrum
$\varepsilon(p)$ and the quasiparticle distribution function
$n_0(p)$ of the FC state \cite{phys_rep}. Because $n_0(p)$
represents the FC state, we have $n_0(p)=1$ at $p\leq p_i$,
$0<n_0(p)<1$ at $p_i<p<p_f$ and $n_0(p)=0$ at $p\geq
p_f$.}}\label{fg4a}
\end{figure}

At temperatures $T\sim 2.3$ K the superconducting-normal phase
transition shown by the solid line in Fig. \ref{fig1} is of the
second order \cite{bianchi,izawa} and entropy $S$ is continuous at
$T_c(B)$. Since $B_{c2}\simeq B_{c0}$, upon the application of
magnetic field $B>B_{c2}$, the HF metal transits to its LFL state
down to lowest temperatures, as seen from Fig. \ref{fig1}. At
$T\to0$ the entropy of the superconducting state $S_{SC}\to 0$ and
the entropy of the NFL state tends to some finite value $S_{NFL}\to
S_0$ \cite{phys_rep94}, since the ground state of systems with FC
is degenerate, and at $T=0$ the quasiparticle distribution function
$n_0({\bf p})$ of flat band is a continuous function in the
interval $[0,1]$,  {in contrast to the Landau FL restriction on
$n_0({\bf p})$ that equals 1 and 0 \cite{lanl1}, see Fig.
\ref{fg4a}}. This property leads to the finite value $S_0$ given by
\begin{equation}
S_0=-\sum_{\bf p} n_0({\bf p})\ln n_0({\bf p})+(1-n_0({\bf
p}))\ln(1- n_0({\bf p})).\label{S*}
\end{equation}
As a result, at sufficiently low temperatures the equality
$S_{SC}(T)=S_{NFL}(T)$ cannot be satisfied \cite{phys_rep,epl06}.
In accordance with the experimental observations
\cite{bianchi,izawa}, the second order phase transition converts to
the first one below some temperature $T_{0}(B)$, being related to
the flat band, induced by FCQPT, and shown by the arrow in Fig.
\ref{fig1} \cite{epl06}. Thus, we conclude that the phase
transition taking place in $\rm CeCoIn_5$ is the symmetry breaking,
since the FC state with a flat band has the special topological
charge different from that of LFL state.
 {In contrast to LFL or marginal
liquids which exhibit the same topological}
 {structure, systems with FC, where
the Fermi surface spreads into a strip, belong to a different
topological class.}  {In case of LFL or marginal liquids the
topological charge takes integer values. The situation is quite
different for systems with FC, where the topological charge becomes
a half-integer, transforming a system with FC into a new class of
Fermi liquids with its own topological structure
\cite{vol,phys_rep94}. The topological FCQPT is of the first order,
since the topological charge is changed abruptly. As a result,
possible fluctuations of order parameters are suppressed
\cite{phys_rep}.}

%\section{Residual resistivity $\rho_0$}

\begin{figure}[!ht]
\begin{center}\vspace*{-0.6cm}
\includegraphics [width=0.47\textwidth]{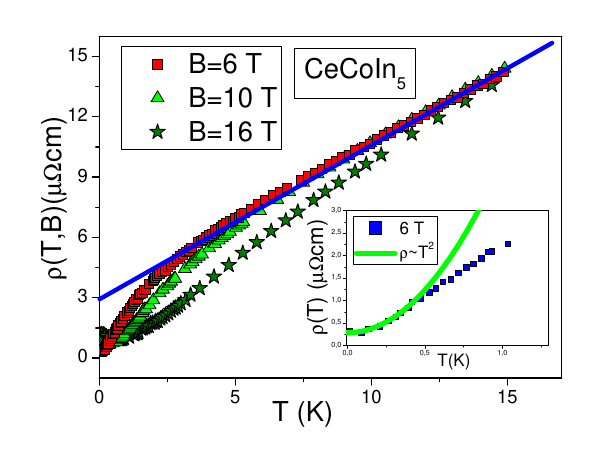}
\end{center}\vspace*{-0.8cm}
\caption{(color online) Resistivity $\rho(T,B)$ obtained in
measurements on $\rm CeCoIn_5$ under the application of magnetic
fields $B$ displayed in the legend \cite{pag1}. The inset exhibits
both the LFL behavior of the resistivity at low temperatures and
the crossover with $1\lesssim n\lesssim 2$ at elevated $T$.}
\label{fig2}
\end{figure}
\begin{figure}[!ht]
\begin{center}\vspace*{-0.8cm}
\includegraphics [width=0.47\textwidth]{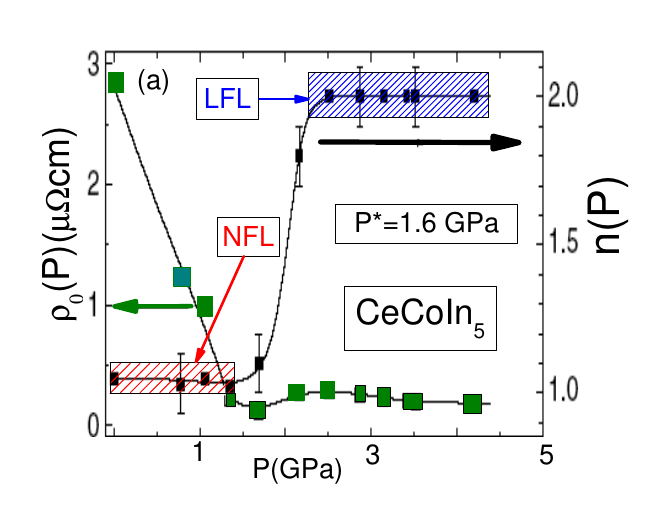}
%\vspace*{-0.8cm}
\includegraphics [width=0.47\textwidth]{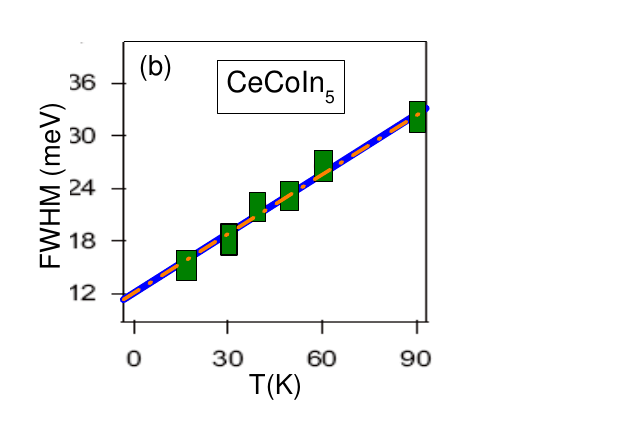}
\end{center}\vspace*{-0.8cm}
\caption{(color online) Panel (a):
 {Values of the residual resistivity
$\rho_0$ (left axis, solid squares) and the index $n$ in the fit
$\rho(T)=\rho_0+AT^n$ (right axis, solid squares) versus pressure
$P$ \cite{sidorov}.} At $P<P^*$ $\rm CeCoIn_5$ is in its NFL state,
at $P>P^*$ is in its LFL one, correspondingly. Both the NFL and LFL
states are shown by the arrows. Panel (b): Temperature dependence
of the full width at half maximum (FWHM) of the single-particle
scattering rate of the main Kondo resonance \cite{chen} is shown by
the solid squares. The line represents the best fit, giving FWHM $=
11.8 + 2.69Tk_B$ meV, where $k_B$ is the Boltzmann constant
\cite{chen}. The dash-dot line is the theory \cite{kz}.}
\label{fig_2}
\end{figure}

The resistivity $\rho(T)$ is of the form
\begin{equation}
\rho(T)=\rho_0+AT^n, \label{res}
\end{equation}
where $\rho_0$ is the residual resistivity and $A$ is a
$T$-independent coefficient. The index $n$ takes the values 1 and
2, respectively, for the NFL and LFL behaviors and $1\lesssim
n\lesssim 2$ in the NFL-LFL transition, see Figs. \ref{fig1},
\ref{fig2} and \ref{fig_2} (a). The residual resistivity $\rho_0$
ordinarily results from impurity scattering \cite{lanl1}. At
$T>T_c$ the zero-field resistivity $\rho(T,B=0)$ varies linearly
with $T$, see Eq. \eqref{res}. At magnetic fields $B\geq B_{c2}$
and low temperatures the resistivity exhibits the LFL behavior,
$\rho(T,B_{c2})\propto T^2$. Experimentally, $\rm CeCoIn_5$ is one
of the purest heavy-fermion metals. As a result, the regime of
electron motion is ballistic. {Thus, under the application of weak
magnetic field $B$ one expect to observe a small positive
contribution $\delta_B\propto B^2$ to $\rho_0$ arising from orbital
motion of electrons induced by the Lorentz force. As seen from
Figs. \ref{fig2} and \ref{fig_2} (a), this is not the case:
specifically $\rho_0\simeq3.0$ ${\rm \mu\Omega cm}$ in the NFL
state, while $\rho_0(B=6\,{\rm T})\simeq 0.3$ ${\rm \mu\Omega cm}$
in the LFL state, see the inset to Fig. \ref{fig2},
\cite{pag1,pag2}, and Fig. \ref{fig_2} (a) \cite{sidorov}. The
observed behavior is defined by the destruction of the FC state by
magnetic field, while the FC state itself creates the additional
residual resistivity \cite{Khod_2020,kz}. We note, that even the
application of strong magnetic fields leads to the negative
magnetoresistivity \cite{pag1}, since the effective mass diminishes
in magnetic fields, $\rho(B)\propto M^*(B)^2\propto B^{-4/3}$,
\cite{phys_rep}. Such a behavior gives the support that $\rm
CeCoIn_5$ is located near the topological FCQPT \cite{phys_rep}.}
Moreover, it is seen from Fig. \ref{fig2} that at elevated
temperatures all the resistivities taken at different fields
$B=6,10,16$ T tend to coincide at the NFL state with
$\rho(T)\propto T$, since the contribution $\delta_B$ is relatively
small.

Another direct experimental confirmation of the change of $\rho_0$
(at the transition from the NFL state to the LFL one) is obtained
in measurements on $\rm CeCoIn_5$ at various pressures $P$
\cite{sidorov}. As seen from Fig. \ref{fig_2} (a),
$\rho_0(P\to0)\to 3.0$ ${\rm \mu\Omega cm}$ with $n=1$, and
decreases by an order of magnitude to a value of about
$\rho_0(P\geq P^*)\to 0.2$ ${\rm \mu\Omega cm}$, with $P^*\simeq
1.6$ GPa and $n=2$ \cite{sidorov}. Note that these values of the
residual resistivity coincide with those shown in Fig. \ref{fig2}.
Obviously, pressure $P$ does not remove impurities from the sample.
Thus, this large decrease in $\rho_0$ is due to a pressure-induced
destruction of the FC state, as seen from the restoration of the
LFL behavior at $P\geq P^*$. Similarly, the resistance $\rho(T,B)$
at fixed $T$ as a function of $B$ diminishes when the system
transits from the NFL behavior to LFL one under the application of
magnetic field (the magnetoresistance becomes negative)
\cite{pag1}. This behavior is consistent with the FC theory
\cite{phys_rep}. Thus, the destruction of the FC state under the
application of both magnetic fields $B$ and pressure $P$ entails a
dramatic suppression of the flat band and its contribution to
$\rho_0$ \cite{kz,Khod_2020}. The same behavior of $\rho_0$ has
been observed on twisted bilayer graphene where strong variation of
$\rho_0$ is seen toward the magic angle  \cite{graph}. In that case
the residual resistivity increases by more than three orders of
magnitude and resembles the behavior of $\rho_0$ shown in Fig.
\ref{fig_2} a  \cite{kz,Khod_2020}.

Now consider the scattering rate $1/\tau$. In the LFL theory
$1/\tau$ is proportional to $T^2$, leading to Eq.~\eqref{res} with
{$n=2$. The NFL behavior comes from the NFL temperature dependence
of $1/\tau$ associated with the presence of FC \cite{kz}. As a
result, the scattering rate becomes $\hbar/\tau\simeq a_1+a_2T$
\cite{kz}.} Here $\hbar$ is Planck's constant, $a_1$ and $a_2$ are
parameters. This result is in good agreement with experimental
facts \cite{tomph,chen}, displayed in Fig. \ref{fig_2} (b). Thus,
the FC theory successfully explains the behavior of the scattering
rate and $\rho(T)\propto T$ \cite{kz}. Taking into account the
above consideration of the resistivity behavior, we conclude that
QCP in $\rm CeCoIn_5$ is not related to the delocalization of
$f$-electrons at QCP, since the delocalization is to lead to the
suppression of the resistivity $\rho(T)$ and does not change the
residual resistivity $\rho_0$. While the suppression of $\rho_0$
takes place when $\rm CeCoIn_5$ transits from the NFL behavior to
LFL one under the application of $B$ or $P$, see Figs. \ref{fig2}
and \ref{fig_2}, and is defined by the destruction of the flat
band, accompanied by the change of the topological charge
\cite{phys_rep94,vol}.

%\section{Thermodynamic properties}

The presence of the flat band manifests itself not only in the
transport properties, but also in the universal scaling behavior of
the thermodynamics properties of HF metals \cite{phys_rep,book_20}.
 {To elucidate this behavior, we
briefly consider the LFL equations defining the effective mass
$M^*(T,B)$. The quasiparticle distribution function $n_\sigma({\bf
p},T)$ has the well known Fermi-Dirac form
\begin{equation}
n_{\sigma}({\bf p},T,B)=\left\{ 1+\exp
\left[\frac{(\varepsilon_{\sigma}({\bf
p},T)-\mu_{\sigma})}T\right]\right\} ^{-1}.\label{HC2}
\end{equation}
Here $\varepsilon_\sigma({\bf p},T)$ is the single-particle energy
spectrum, $\mu$ is the chemical potential, being spin $\sigma$
dependent due to the Zeeman splitting $\mu_{\sigma}=\mu\pm \mu_BB$,
with $\mu_B$ is the Bohr magneton. The spectrum
$\varepsilon_{\sigma}({\bf p},T)$ is obtained from the system
energy $E[n_{\sigma}({\bf p},T)]$,
\begin{equation} \label{rac}
\varepsilon_{\sigma}({\bf p},T)= \frac{\delta E[n({\bf p})]}{\delta
n_{\sigma}}.
\end{equation}
The effective mass $M^*(T,B)$ is given by
\cite{lanl1,phys_rep,book_20}.
 {\begin{equation}
\frac{1}{M^*(T,B)}=\frac{1}{m}+\sum_{\sigma}\int\frac{{\bf p}_F{\bf
p}}{p_F^3}F ({\bf p_F},{\bf p})\frac{\partial n_{\sigma} ({\bf
p})}{\partial{p}}\frac{d{\bf p}}{(2\pi)^3}. \label{HC1}
\end{equation}}
Here, $m$ is the free electron mass, $F({\bf p_F},{\bf p})$ is the
Landau interaction, depending on momentum $\bf p$ and the Fermi
momentum $\bf p_F$}. The main goal of the quasiparticle interaction
$F({\bf p},{\bf p}_1)$ is to place the system at the topological
FCQPT \cite{phys_rep,book_20}. Thus, the universal scaling behavior
of HF metals can be explained, for it becomes independent of
interactions near the formation of flat bands
\cite{phys_rep,book_20}. Equations \eqref{HC2}, \eqref{rac} and
\eqref{HC1} constitute the closed set that allows one to find
$\varepsilon_\sigma({\bf p},T)$ and $n_{\sigma}({\bf p},T)$. In
this case the effective mass is determined by the expression
$p_F/M^*=\partial\varepsilon(p)/\partial p|_{p=p_F}$. At the
topological FCQPT point, the analytical solutions of
Eq.~\eqref{HC1} are possible, see e.g. \cite{phys_rep}. In contrast
to the LFL theory with $M^*$ being an approximatively constant
parameter, here at zero magnetic field, $M^*$, exhibiting the NFL
behavior, becomes temperature dependent. This feature represents
the vivid deviation from the LFL picture, determining the NFL
regime
\begin{equation}
M^*(T)\simeq a_TT^{-2/3},\label{MTT}
\end{equation}
where $a_T$ is fitting parameter. At raising temperatures and under
the application of magnetic filed $B$, the system undergoes a
transition to the LFL behavior, so that the effective mass becomes
approximately independent of $T$ and $\rho(T)\propto T^2$, see
Figs. \ref{fig1}, \ref{fig2}.
\begin{equation}
 {M^*(B)\simeq
a_B(B-B_{c0})^{-2/3}\label{MBB}}
\end{equation}
where $a_B$ is fitting parameter.  We recall that in case of $\rm
CeCoIn_5$ the critical field $B_{c0}\simeq $ 5.0 T. In some cases
as in the HF metal $\rm CeRu_2Si_2$, $B_{c0}=0$ \cite{takah}, and
in $\rm YbRh_2Si_2$, $B_{c0}\simeq 0.07$ T \cite{yb}. {It is seen
from Eq. \eqref{MTT} that a HF metal located near FCQPT
demonstrates the typical NFL behavior related to the strong
dependency of $M^*$ on $T$, so that $M^*(T\to0)\to\infty$. It
follows from Eq. \eqref{MBB} that the application of $B$ makes HF
metal transits to the LFL state with $M^*(T)$ exhibiting the LFL
behavior \cite{phys_rep}. These both prominent features are not
observed in the case of the Landau Fermi liquid where $M^*$ is of
the same order of magnitude as the bare mass \cite{lanl1}.}

Now we turn to the universal scaling behavior of $M^*$. The
introduction of "internal" (or natural) scales greatly clarifies
the problem of the scaling behavior.  We first observe that near
the topological FCQPT, both the effective mass $M^*(B,T)$ and the
solution of Eq.~\eqref{HC1} have the maximum $M^*_M$ at a
temperature $T_{M}\propto B$ \cite{phys_rep}.
\begin{figure}
\begin{center}
\vspace*{-0.5cm}
\includegraphics [width=0.5\textwidth]{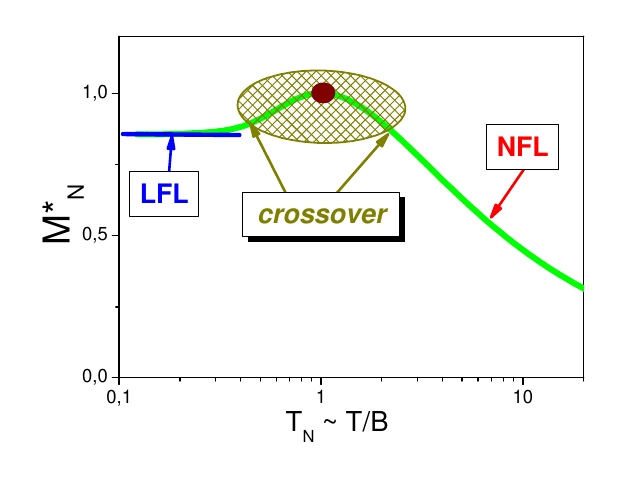}
\vspace*{-1.5cm}
\end{center}
\caption{(color online) Universal scaling behavior of the
thermodynamic properties of HF metals defined by the normalized
effective mass $M^*_N$. As follows from Eq. \eqref{TMB}, under the
application of magnetic field $T_N\propto T/(B-B_{c0})$. Solid
curve depicts $M^*_N$ versus normalized temperature $T_N$. It is
seen that at finite $T_N<1$ the LFL behavior takes place. At
$T_N\sim 1$ the system enters crossover state, and at elevating
temperatures demonstrates the NFL behavior.}\label{fig9}
\end{figure}

It is convenient to measure the effective mass and temperature in
the units $M^*_M$ and $T_{M}$ respectively \cite{phys_rep}, and we
arrive at normalized effective mass $M^*_N=M^*/M^*_M$ and
temperature $T_N=T/T_{M}$. Using the internal scales $M_N^*$ and
$T_N$ allows us to reveal the universal scaling behavior, see Fig.
\ref{fig9}, since $M_N^*$ and $T_N$ do not depend on the specific
properties of HF metal, while $M_M^*$ and $T_M$ do
\cite{phys_rep,book_20}. Near the topological FCQPT, the dependence
$M^*_N(T_N)$ can be approximated by the universal interpolating
function that describes the transition from LFL to NFL states,
given by Eqs. \eqref{MTT} and \eqref{MBB}, and represents the
universal scaling behavior of $M^*_N$ \cite{phys_rep,book_20}
\begin{equation}M^*_N(y)\approx c_0\frac{1+c_1y^2}{1+c_2y^{8/3}}.
\label{UN2}
\end{equation}
Here $y=T_N=T/T_{M}$ and $c_0=(1+c_2)/(1+c_1)$, where $c_1$ and
$c_2$ are fitting parameters. The magnetic field $B$ enters
Eq.~\eqref{HC1} only in the combination $\mu_BB/T$, making
$T_{M}\sim \mu_BB$. As a result, we obtain from Eq.~\eqref{UN2}
that
\begin{equation}
 {\label{TMB2} T_M\simeq
a_1\mu_B(B-B_{co})},
\end{equation}
where $a_1$ is a dimensionless fitting parameter, see e.g.
\cite{phf}. In this case, the variable $y$ becomes
$y=T/T_{M}\propto T/\mu_B(B-B_{c0})$. Equation \eqref{TMB2} permits
to assert that Eq.~\eqref{UN2} gives the universal scaling
properties of the effective mass $M^*_N(y)$. Thus, the
thermodynamic functions measured at different field form a single
function $M^*_N(T_N=y)$ \cite{phys_rep}. Since $T$ and $B$ enter
symmetrically in Eq.~\eqref{UN2}, it also manifests the scaling
behavior of $M^*_{N}(B,T)$ as a function of $T/B$:
\begin{equation}
 {\label{TMB} T_N=\frac{T}{T_M}=
\frac{T}{a_1\mu_B(B-B_{co})}\propto \frac{T}{(B-B_{co})}\propto
\frac{(B-B_{co})}{T}.}
\end{equation}
This universal scaling exhibited by $M_N$ is also shown in Fig.
\ref{fig9}. We note that Eqs. \eqref{UN2} and \eqref{TMB} allow one
to describe the universal scaling behavior of HF metals, see e.g.
\cite{phys_rep,book_20}.

\begin{figure}[!ht]
\begin{center}\vspace*{-0.5cm}
\includegraphics [width=0.47\textwidth]{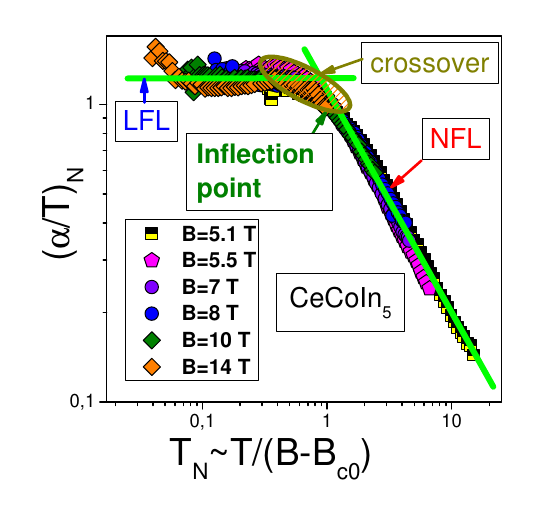}
\end{center}\vspace*{-0.8cm}
\caption{(color online) Normalized low temperature thermal
expansion coefficient $(\alpha/T)_N$ vs $T_N$ at different $B$
shown in the legend \cite{kz}. The data \cite{steg,tompson} and $T$
were normalized by the values of $\alpha/T$ and by the temperature
$T_{inf}\simeq T_{cross}$, respectively, at the inflection point
shown by the arrow. The horizontal solid line shows the LFL
behavior, $(\alpha/T)={\rm const}$. The other solid line depicts
the NFL behavior $(\alpha/T)\propto 1/T_N$.} \label{fig3}
\end{figure}
Since the properties of HF metals are formed by the topological
FCQPT inducing flat bands, see e.g. \cite{phys_rep,book_20}, the
behavior of the dimensionless thermal expansion coefficient of $\rm
CeCoIn_5$, treated as a function of the dimensionless temperature
$T_N$, is also universal. Figure \ref{fig3} shows that all the
normalized data extracted from measurements on $\rm CeCoIn_5$
\cite{steg,tompson} collapse onto a single scaling curve \cite{kz}.
As seen from the Fig. \ref{fig3}, the dimensionless coefficient
$(\alpha(T,B)/T)_N$, treated as a function of $T_N$, at $T_N<1$
shows a constant value, as depicted by the line, implying $\rm
CeCoIn_5$ exhibits the LFL behavior. At $T_N\simeq 1$ the system
enters the narrow crossover region. At $T_N>1$, the NFL behavior
takes place, and $(\alpha/T)_N\propto 1/T_N$. From this observation
we conclude that the essential features of the experimental $T-B$
phase diagram of $\rm CeCoIn_5$ are well represented by Fig.
\ref{fig1}.

%\section{Asymmetric tunneling differential conductivity}\label{tun}

\begin{figure} [! ht]
\begin{center}\vspace*{-0.8cm}
\includegraphics [width=0.47\textwidth] {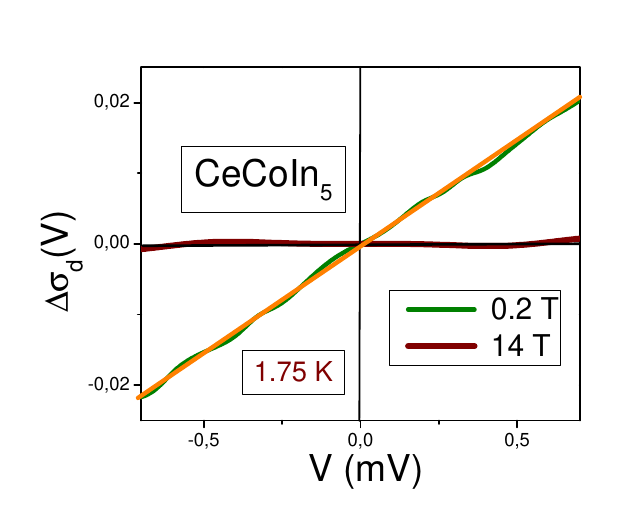}
\end{center}\vspace*{-0.8cm}
\caption{(color online) Asymmetric part $\Delta\sigma_d(V)$ of the
tunneling differential conductivity measured on $\rm CeCoIn_5$ and
extracted from the experimental data \cite{park1}. Linear
dependence of $\Delta\sigma_d$ is shown by the straight line. The
asymmetric part disappears at $B=14$ T and $T=1.75$ K, with
$B_{c0}\simeq 5$ T.} \label{Fig4}
\end{figure}

We now consider the differential tunneling conductivity
$\sigma_d(V)$ between a HF metal and a simple metallic point. At
low temperatures $\sigma_d(V)$ can be noticeably asymmetrical with
respect to the change of voltage bias $V$ that makes the
asymmetrical part $\Delta\sigma_d(V)=\sigma_d(V)-\sigma_d(-V)$
finite. The asymmetry can be observed in experiments on HF metals
whose electronic system has undergone the topological FCQPT, while
the application of magnetic field causes the system to exhibit the
LFL behavior, and eliminates the asymmetry, as has been predicted
\cite{tun,symm}. Such a behavior has been observed in measurements
on the HF metal $\rm CeCoIn_5$ \cite{park1}, displayed in
Fig.~\ref{Fig4}. The data exposed in Fig.~\ref{Fig4} have been
extracted from \cite{park1}, Fig. S17, and demonstrate the sample
independence of the results. {It is seen from Fig.~\ref{Fig4}, that
$\Delta\sigma_d(V)$ is a linear function of $V$, and vanishes under
the application of magnetic field that restores the LFL behavior
\cite{phys_rep,tun,ph_scr,atom}.}

\begin{figure}[!ht]
\begin{center}
\vspace*{-0.3cm}
\includegraphics[width=0.47\textwidth] {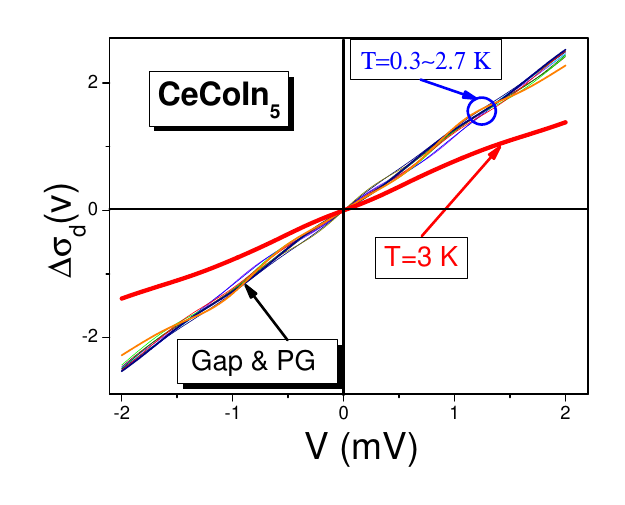}
\end{center}
\vspace*{-0.8cm} \caption{(color online) {The asymmetric part of
tunneling conductivity $\Delta\sigma_d(V)$ in $\rm CeCoIn_5$.
$\Delta\sigma_d(V)$ is extracted from the experimental data
\cite{steg2014}.} At $2.3\leq T\leq 2.7$ K $\rm CeCoIn_5$ in its
pseudogap (PG) and at $T\leq 2.3$ in its superconducting states
with $T_c=2.3$ \cite{steg2014}. At $T\leq 2.7$ K, as it is shown by
both the ring and the arrow, $\Delta\sigma_d(V)$ is temperature
independent \cite{phys_rep,shag18,pla07}.} \label{fig7}
\end{figure}

{Now consider how the presence of $\Delta\sigma_d(V)$ signals the
violation of $\mathcal{T}$ and $\mathcal{C}$ symmetries
\cite{ph_scr,symm}. Suppose that we have a contact between HF and
ordinary metals. Let us initially have the electronic current
directed from HF metal to an ordinary one. Upon applying voltage
$V$ to the contact, we also change the electron charge $-e$ to $+e$
which alters immediately the current direction. Consequently, one
obtains exactly the above electronic current under the voltage sign
change $V \to -V$. As a result, the differential conductivity
acquires the same asymmetric part $\Delta\sigma_d(V)$, as it is
seen from Fig. \ref{Fig4}. If $\mathcal{C}$ were conserved, the
asymmetrical part $\Delta\sigma_d(V)=0$. Thus, we conclude that
because of finite value of $\Delta\sigma_d(V)$ the $\mathcal{C}$
symmetry is broken. Also, the substitution $t \to -t$ for constant
charge generates the change of current direction only, where $t$ is
time.} {Since, the latter direction change can be accomplished by
$V\to -V$ also, it is clear that the time inversion symmetry is
violated if $\Delta\sigma_d(V)$ is finite.} {Hence, both
$\mathcal{C}$ and $\mathcal{T}$ symmetries are violated, provided
that $\Delta\sigma_d(V)\neq 0$ emerges. Concurrently, the
simultaneous transform $e\to-e$ and $t\to-t$ does not change
anything, which means that combined $\mathcal{C,T}$ symmetry is
preserved. It worth noting that in the present case the symmetry
$\mathcal{P}$ with respect to coordinates sign flip, that is the
parity $\mathcal{P}$ is not violated, so that the combined general
$\mathcal{C,P,T}$ symmetry is kept intact \cite{symm,ph_scr}.} It
is well known that both $\mathcal{C}$ and $\mathcal{T}$ symmetries
are conserve for the systems of fermions, described by Landau
theory. This implies that for these systems like ordinary metals
$\sigma_d(V)$ is a symmetric function of its variable $V$, so that
conductivity asymmetry $\Delta\sigma_d(V)$ is not observed in them
at low $T$. {As a result, in magnetic fields both the FC state and
the asymmetry vanish, as soon as the LFL behavior is restored, see
e.g. \cite{phys_rep,tun,ph_scr,symm}. We note that the application
of magnetic field itself violates the $\mathcal{T}$ symmetry, but
this violation is rather weak to be observed in measurement of
tunneling conductivity.} Thus, $\Delta\sigma_d(V)\neq 0$ signals
the presence of FC and the corresponding flat band that make the
violation of both $\mathcal{T}$ and $\mathcal{C}$ finite
\cite{ph_scr,symm}. It is seen from Fig. \ref{fig7} that $\rm
CeCoIn_5$ in its both the superconducting state and pseudogap one
exhibits asymmetrical tunneling conductivity $\Delta\sigma_d(V)\neq
0$. The experimental data \cite{steg2014} displayed in Fig.
\ref{fig7} are in good agreement with the data obtained in Ref.
\cite{park}, as it is shown in Ref. \cite{shag18}. These facts
confirm that beyond $B_{c0}$ the FC state takes place, promoting
both the superconducting and the corresponding pseudogap (PG)
states \cite{shag18}. It is seen from Fig. \ref{fig7}, that at
$T\leq 2.7$ K $\Delta\sigma_d(V)$ is temperature independent, while
$\rm CeCoIn_5$ is in its superconducting and pseudogap states. Such
a behavior is the intrinsic feature of HF metals located near the
topological FCQPT \citep{phys_rep,pla07,atom}. We note that the
violation of $\mathcal{C}$ in strongly correlated Fermi systems was
experimentally observed \cite{prx}, as it was predicted
\cite{tun,phys_rep94,phys_rep,symm}. The interaction-driven
violation of $\mathcal{C}$ was observed in twisted bilayer graphene
\cite{step,paul}. Both this observation and the violation of
$\mathcal{T}$ in graphene support that the flat band of graphene is
formed due to the topological FCQPT \cite{ph_scr}.

It is seen from the phase diagram \ref{fig1} that at raising
magnetic fields $B$ and low temperatures $T$ the HF metal $\rm
CeCoIn_5$ exhibits the LFL behavior and the symmetries are restored
\cite{ph_scr,symm}.  {The same behavior is exhibited by the HF
metals $\rm YbRh_2Si_2$ \cite{seiro}, $\rm YbCu_{5-x}Al_x$ (for
x=1.5 \cite{pris}) and graphene \cite{cao}, and explained within
the framework of the FC theory \cite{phys_rep,shag18}. We note that
the asymmetrical part $\Delta\sigma_d(V)\neq 0$ is observed in the
normal state of HF metals without a change at $T\simeq T_c$, and
diminishing when the the pseudogap vanishes, see Fig. \ref{fig7}
\cite{pris,phys_rep,book_20}. Under the application of magnetic
field this asymmetry vanishes, thus possible in gap states in the
superconducting state do not noticeably contribute to
$\Delta\sigma_d(V)$.} Again, we conclude that the phase transition
in $\rm CeCoIn_5$ is represented by the topological FCQPT and
accompanied by the symmetry violation, as it does in the mentioned
above HF metals and graphene.

%\section{Hall effect}

We are now in a position to consider in $\rm CeCoIn_5$ a possible
jump in the Hall coefficient at $B\to B_{c0}$ at the zero
temperature limit. Measurements of the Hall resistivity $R_H(T,B)$
in external magnetic fields $B$ versus $T$ have revealed a diverse
low-temperature behavior of this basic property in HF metals,
ranging from the LFL behavior to the challenging NFL one
\cite{hall22,pash,phys_rep}.

It is instructive to compare the behavior of the Hall coefficient
in archetypical HF metals $\rm CeCoIn_5$ and $\rm YbRh_2Si_2$. At
$T\to0$ in $\rm YbRh_2Si_2$ the application of the critical
magnetic field $B_{c0}$ suppressing the antiferromagnetic phase
(with the Fermi momentum $p_{AF}\simeq p_F$) restores the LFL
behavior with the Fermi momentum $p_f>p_F$, making an abrupt change
in the Hall coefficient $R_H(T\to0,B\to B_{c0})$, as a function of
$B$ \cite{pash,phys_rep}. At low temperatures and $B<B_{c0}$, the
ground state energy of the antiferromagnetic phase is lower than
that of the heavy LFL, while at $B>B_{c0}$ the opposite case takes
place, and the LFL state wins the competition. At $B=B_{c0}$ both
the antiferromagnetic and LFL states have the same ground state
energy. Thus, at $T=0$ and $B=B_{c0}$, the infinitesimally small
change in the magnetic field $B$ leads to the finite jump in the
Fermi momentum. In response to this change, the Hall coefficient
$R_H(B)$ undergoes the corresponding sudden jump
\cite{pash,phys_rep,kir,tau}. In the case of $\rm CeCoIn_5$, one
can hardly expect to observe such a jump, since QCP is hidden under
the superconducting dome, see the phase diagram \ref{fig1}. As a
result, one can only observe $R_H(B>B_{c0})$. Measurements of the
Hall coefficient on $\rm CeCoIn_5$ do not show the possible jump,
while the observed temperature dependence of $R_H(T,B)$
\cite{hall22} can be explained within the framework of the FC
theory \cite{khodh}. The absence of the jump in the Hall
coefficient makes the illusion that the quantum phase transition in
$\rm CeCoIn_5$ is not accompanied by the violation of symmetry, but
the study of the $\mathcal{T}$ and $\mathcal{C}$ symmetries and the
change of the residual resistivity $\rho_0$ clearly signals that
the violation of the symmetry does take place.

%\section{Summary}
 {In summary: We have delineated the
quantum phase transition in $\rm CeCoIn_5$, and explained a number
of experimental data collected in measurements on $\rm CeCoIn_5$.
The quantum phase transition  in $\rm CeCoIn_5$ is accompanied by
the symmetry breaking and represented by the topological FCQPT that
connects two Fermi surfaces of different topological charges and
forms flat bands.}

This work was supported by U.S. DOE, Division of Chemical Sciences,
Office of Basic Energy Sciences, Office of Energy Research, AFOSR.


\begin{thebibliography}{99}

\bibitem{ks} V.A. Khodel and V.R. Shaginyan,
JETP Lett. {\bf 51}, 553 (1990).

\bibitem{phys_rep94} V.A. Khodel, V.R. Shaginyan, and V.V.
Khodel, Phys. Rep. {\bf 249}, 1 (1994).

\bibitem{hall22} N. Maksimovic,
D.H. Eilbott, T. Cookmeyer, F. Wan, J. Rusz {\it et al.}, Science
{\bf 375}, 76 (2022).

\bibitem{voj} M. Vojta, Rep. Prog. Phys. {\bf 66}, 2069 (2003).

\bibitem{phys_rep} V.R. Shaginyan, M.Ya. Amusia, A.Z. Msezane,
and K.G. Popov, Phys. Rep. {\bf 492}, 31 (2010).

\bibitem{book_20} M.Ya. Amusia and V.R. Shaginyan,
{\it Strongly Correlated Fermi Systems: A New State of Matter},
Springer Tracts in Modern Physics Vol. {\bf 283} (Springer Nature
Switzerland AG, Cham, 2020).

\bibitem{Khod_2020} V.A. Khodel, J.W. Clark, and M.V. Zverev,
\prb {\bf 102}, 201108(R) (2020).

\bibitem{vol} G.E. Volovik, JETP Lett. {\bf 53}, 222 (1991).

\bibitem{kir} {S. Kirchner, S. Paschen,
Q. Chen, S. Wirth, D. Feng {\it et al.}, Rev. Mod. Phys. {\bf 92},
011002 (2020).}

\bibitem{tau} {M. Taupin and S. Paschen, Crystals {\bf 12}, 251 (2022).}

\bibitem{catal}  {N. Regnault, Y. Xu, M.-R. Li, D.-Sh. Ma, M.
Jovanovic {\it et al.}, Nature {\bf 603}, 824 (2022).}

\bibitem{pash} S. Paschen, T.L\"uhmann, S. Wirth, P. Gegenwart,
O. Trovarelli {\it et al.}, Nature {\bf 432}, 881 (2004).

\bibitem{lanl1}  E.M. Lifshitz and L.P. Pitaevskii,
{\it Statistical Physics,} Part 2, Butterworth-Heinemann (1999).

\bibitem{ph_scr} V.R. Shaginyan, A.Z. Msezane,
V.A. Stephanovich, G.S. Japaridze, and E.V. Kirichenko, Phys. Scr.
{\bf 94}, 065801 (2019).

\bibitem{symm} V.R. Shaginyan, A.Z.  Msezane, G.S. Japaridze, and V.A.
Stephanovich, Symmetry {\bf 12}, 1596 (2020).

\bibitem{atom} V.R. Shaginyan, A.Z. Msezane, and G.S. Japaridze,
Atoms {\bf 10}, 67 (2022).

\bibitem{chen} Q.Y. Chen,
D.F. Xu, X.H. Niu, J.Jiang, R. Peng {\it et al.}, \prb {\bf 96},
045107 (2017).

\bibitem{qcp11} L. Howald, G. Seyfarth, G. Knebel, G. Lapertot1, D. Aoki,
and J.P. Brison, J. Phys. Soc. Jpn., {\bf 80}, 024710 (2011).

\bibitem{pag1} J. Paglione, M. A. Tanatar, D. G. Hawthorn,
E. Boaknin, R. W. Hill {\it et al.}, \prl {\bf 91}, 246405 (2003).

\bibitem{pag2} J. Paglione, M. A. Tanatar,
D. G. Hawthorn, F. Ronning, R. W. Hill {\it et al.}, \prl {\bf 97},
106606 (2006).

\bibitem{bianchi} A. Bianchi, R. Movshovich, N. Oeschler,
P. Gegenwart, F. Steglich {\it et al.}, \prl {\bf 89}, 137002
(2002).

\bibitem{sc08} {M. Kenzelmann,
Th. Str\"assle, C. Niedermayer, M. Sigrist, B. Padmanabhan {\it et
al.}, Science {\bf 321}, 1652 (2008).}

\bibitem{prl10} {M. Kenzelmann,
S. Gerber, N. Egetenmeyer, J.L. Gavilano, Th. Str\"assle {\it et
al.}, \prl {\bf 104}, 127001 (2010).}

\bibitem{ronn} F. Ronning, C. Capan, E.D. Bauer,
J.D. Thompson, J.L. Sarrao, and R. Movshovich, \prb {\bf 73},
064519 (2006).

\bibitem{izawa} K. Izawa, H. Yamaguchi,
Y. Matsuda, H. Shishido, R. Settai, and Y. Onuki, \prl {\bf 87},
057002 (2001).

\bibitem{epl06} V.R. Shaginyan, A.Z. Msezane, V.A. Stephanovich, and E.V.
Kirichenko, Europhys. Lett. {\bf 76}, 898 (2006).

\bibitem{sidorov}  V.A. Sidorov, M. Nicklas, P.G. Pagliuso,
J.L. Sarrao, Y. Bang {\it et al.}, \prl {\bf 89}, 157004 (2002).

\bibitem{kz} V.R. Shaginyan, A.Z. Msezane, K.G. Popov, J.W. Clark, M.V.
Zverev, and V.A. Khodel, \prb {\bf 86}, 085147 (2012).

\bibitem{graph} H. Polshyn, M. Yankowitz, S. Chen, Y. Zhang, K. Watanabe
{\it et al.}, Nat. Phys. {\bf 15}, 1011 (2019).

\bibitem{tomph} P. Aynajian, E. Neto, A. Gyenis,
R.E. Baumbach, J.D. Thompson {\it et al.}, Nature {\bf 486}, 201
(2012).

\bibitem{takah} D. Takahashi, S. Abe, H. Mizuno, D.A. Tayurskii,
K. Matsumoto {\it et al.}, Phys. Rev. B {\bf 67}, 180407 (2003).

\bibitem{yb} P. Gegenwart, Y. Tokiwa, T. Westerkamp,
F. Weickert, J. Custers {\it et al.}, New J. Phys. {\bf 8}, 171
(2006).

\bibitem{phf} V.R. Shaginyan, A.Z. Msezane, G.S. Japaridze, S.A. Artamonov,
and Y.S. Leevik, Materials {\bf 15}, 3901 (2022).

\bibitem{steg} J. G. Donath, F.
Steglich, E. D. Bauer, J. L. Sarrao, and P.~Gegenwart, \prl {\bf
100}, 136401 (2008).

\bibitem {tompson} S. Zaum, K. Grube, R. Sch\"afer,
E. D. Bauer, J. D. Thompson, and H. v. L\"ohneysen \prl {\bf 106},
087003 (2011).

\bibitem{tun}  V.R. Shaginyan, JETP Lett. {\bf 81}, 222 (2005).

\bibitem{park1} K. Shrestha, S. Zhang, L.H. Greene,
Y. Lai, R.E. Baumbach {\it et al.}, \prb {\bf 103}, 224515 (2021).

\bibitem{steg2014} S. Wirth, Y. Prots, M. Wedel, S. Ernst,
S. Kirchner {\it et al.}, J. Phys. Soc. Japan {\bf 83}, 061009
(2014).

\bibitem{park} W.K. Park, L.H. Greene, J.L. Sarrao, and J.D.
Thompson, Phys. Rev. B {\bf 72}, 052509 (2005).

\bibitem{shag18} V.R. Shaginyan, A.Z. Msezane, G.S. Japaridze,
V.A. Stephanovich, and Y.S. Leevik, JETP Lett. {\bf 108}, 335
(2018).

\bibitem{pla07} V.R. Shaginyan and K.G. Popov,
Phys.  Lett.  A {\bf 361}, 406 (2007).

\bibitem{prx} A. Gourgout, G. Grissonnanche, F. Lalibert\`e,
A. Ataei, L. Chen {\it et al.}, Phys. Rev. X {\bf 12}, 011037
(2022).

\bibitem{step} P. Stepanov, Nat. Phys. {\bf 18}, 608 (2022).

\bibitem{paul}  {A.K. Paul, A. Ghosh, S. Chakraborty, {\it et al.,}
Nature Physics {\bf 18}, 691 (2022).}

\bibitem{seiro} S. Seiro, L. Jiao, S. Kirchner, S. Hartmann,
S. Friedemann {\it et al.}, Nat. Commun. {\bf 9}, 3324 (2018).

\bibitem{pris} G. Prist\`a\~s, M. Reiffers, E. Bauer, A.G.M. Jansen,
and D.K. Maude, Phys. Rev. B {\bf 78}, 235108 (2008).

\bibitem{cao} Y. Cao, V. Fatemi, S. Fang, K. Watanabe,
T. Taniguchi {\it et al.}, Nature (London, U.K.) {\bf 556}, 43
(2018).

\bibitem{khodh} V.A. Khodel, M.V. Zverev, and J. W. Clark,
JETP Lett. {\bf 81}, 315 (2005).

\end{thebibliography}
\end{document}